\documentclass[twocolumn,american,english,aps, pra,eqsecnum]{revtex4-2}
\usepackage[T1]{fontenc}
\usepackage[latin9]{inputenc}
\usepackage{color}
\usepackage{bm}
\usepackage{amsmath}
\usepackage{amssymb}
\usepackage{graphicx}
\usepackage{babel}
\begin{document}
\title{Simulating complex networks in phase space: Gaussian boson sampling}
\author{Peter D. Drummond, Bogdan Opanchuk, A. Dellios and M. D. Reid}
\affiliation{Centre for Quantum Science and Technology Theory, Swinburne University
of Technology, Melbourne 3122, Australia}
\begin{abstract}
We show how phase-space simulations of quantum states in a linear
photonic network permit the verification of measurable probabilities
and entanglement. We compare our predictions with recent Gaussian
boson sampling experiments of Zhong \emph{et al}. These use squeezed
inputs and efficient ``on-off'' detectors, with up to $76$-th order
measured coincidence counts in the data. We introduce a general definition
of grouped ``on-off'' detection probabilities for this purpose.
The positive-$P$ phase-space method is used to compute any grouped
or marginal click probabilities. Additional decoherence is included
to obtain agreement between theory and experiment. The only limitation
in estimating grouped probabilities is the computational sampling
error, which is similar in magnitude to the experimental sampling
error. The results obtained and graphed here are from first-order
up to 16 000-th order grouped count probabilities. However, any order
between these is also computable. We extend these results to include
grouped probabilities with multidimensional outcomes that have a polynomial
number of points. We also analyze quadrature detection experiments,
and show how to simulate genuine multipartite entanglement using Wigner
phase-space methods. 
\end{abstract}
\maketitle

\section{Introduction\label{sec:Introduction}}

Bosonic quantum networks are increasingly useful in quantum technology
and quantum computing applications \citep{Scheel2005}. Linear networks
driven either by nonclassical number state \foreignlanguage{american}{\citep{Aaronson2011,broome2013photonic,crespi2013integrated,tillmann2013experimental,spring2013boson,spagnolo2014experimental,Crespi:2014}}
or Gaussian inputs \citep{lund2014boson,Hamilton2017PhysRevLett.119.170501,quesada2018gaussian,kruse2019detailed}
for boson sampling are becoming widely available. The squeezed-state
interferometer, which is a two-mode linear network, is being employed
to enhance gravitational-wave detection sensitivity \citep{Caves_PRD1981,McCuller2020PhysRevLett.124.171102}.
More complex photonic networks are under development, both as novel
interferometers \citep{Motes2015_PRL114,Su2017Multiphoton} and as
test-beds for multipartite entanglement \citep{Chen2014PhysRevLett.112.120505,roslund2014wavelength,yoshikawa2016invited}.
Other examples include the Ising machine, used to solve large \emph{NP}-hard
optimization problems \citep{Marandi_CIM_Nature2014,inagaki2016coherent,yamamoto2017coherent}.

A dramatic increase in scale of a boson sampling quantum network has
recently been achieved. Zhong \emph{et al.} \citep{zhong2020quantum}
implemented a $100$-mode Gaussian boson sampler (GBS) with squeezed
inputs, and detected the output photon counts, whose distribution
is called the 'Torontonian'. They measured up to $76$-th order coincidence
counts in the outputs, which they estimated to take $0.6$ billion
years to simulate conventionally on the world's fastest current supercomputer.
This has led to reports of quantum supremacy by Zhong \emph{et al}.:
that is, a quantum device implementing a computational task that is
not classically feasible \citep{Bremner2016PhysRevLett.117.080501,boixo2018characterizing}.
Similar reports have been made using quantum logic gates \citep{arute2019quantum}.

There is an ongoing debate on how to rigorously validate such technology
\citep{shepherd2009temporally,Hangleiter2019PhysRevLett.122.210502}.
Validation based on low-order correlations \citep{phillips2019benchmarking}
or direct classical simulation may be susceptible to mock-ups. However,
Gaussian inputs to linear networks \citep{lund2014boson,brod2019photonic}
have a non-computable discrete count probability for large mode number.
The output distribution is a Hafnian \citep{Hamilton2017PhysRevLett.119.170501}
for photon-number-resolving detectors. In the ``on-off'' or saturable
detector case, it is the Torontonian \citep{quesada2018gaussian}.
Quesada \emph{et al.} \citep{quesada2018gaussian,quesada2020exact}
explained that classical evaluation of the ideal ``Torontonian''
distribution is exponentially complex, making it nearly impossible
for more than $50$ modes \citep{li2020benchmarking}. 

Since a direct simulation of the output correlations at large $M$
cannot be achieved in less than exponential time, another approach
is needed to compare theory with experiment in a reasonable timeframe.
Theoretical benchmarks are essential, both to know what output is
expected, and to understand any other physics. 

The objective of this paper is to do simply this: to investigate whether
the theory agrees with experiment. We show how one may in part solve
this problem, by simulating Gaussian boson linear networks in quantum
phase space. This provides a way to verify the output quantum correlations
and marginal probabilities, both in the idealized case and with other
known physics included. It is important to note that we do not simulate
the experiment directly using discrete photon counts \citep{quesada2018gaussian,quesada2020exact}.
Rather, we verify observable, grouped probabilities by averaging over
many random trajectories in phase space \citep{opanchuk2018simulating},
which have the same correlations and marginal probabilities. 

By contrast, an explicit simulation with discrete counts is feasible,
but is currently limited to either small networks or large decoherence
\citep{qi2020regimes}. Our methods provide a way to certify the measurable
probabilities of experimental outputs, even for systems much larger
than the current experimental ones.  The phase space approach uses
the correctness of quantum mechanics under a change of basis state
to develop an alternative, powerful algorithm for efficient verification
of any grouped count distribution. We have computed all of the grouped
probabilities measured by Zhong \emph{et al.} \citep{zhong2020quantum},
including $76$-th order coincidence counts, as well as other marginal
probabilities in their data. We calculate these from the model of
the apparatus and experimental parameters measured by the experimentalists. 

Differences between theory and experiment appear to be caused by input
pulse decoherence effects. Our phase-space simulations agree much
better with experimental count distributions \citep{zhong2020quantum}
after including these decoherence effects, and are scalable. Chi-squared
tests were carried out \citep{pearson1911probability,knuth2014art},
which show an improvement of three orders of magnitude for theoretical
agreement with experiment.

Results are also obtained for up to $16\,000$ mode devices. Computational
times and sampling errors are comparable to those in experiments for
the same number of samples, making them useful for validation of GBS
experiments. We computed marginal, low-order probabilities, as well
as higher order, measurable grouped probabilities. Our methods can
also be extended if necessary to include other known physics, including
multiple frequency modes, dispersion, nonlinearity, decoherence and
Raman/Brillouin scattering \citep{corney2008simulations}, allowing
better understanding of these experiments.

Boson sampling outputs are exponentially hard to generate numerically.
Here, we demonstrate how measurable grouped probabilities can be verified.
A key feature of quantum mechanics which allows this in GBS is that
binary ``click'' measurement operators are multimode projection
operators, whose averages are probabilities. The positive-$P$ phase-space
representation allows one to calculate these probabilities using efficient
phase-space sampling methods, even though a direct sampling of the
click distribution is exponentially hard. Since methods used to detect
eavesdroppers involve measurement of moments or probabilities, this
also suggests relevance to cryptographic steganography, in which messages
could be hidden in the random outputs. We further demonstrate, using
simulations in a different phase-space, the Wigner representation,
how to certify the genuine $M$-partite entanglement of all $M=100$
nodes of a Gaussian network.

\section{\label{sec:Gaussian-bosonic-networks}Gaussian bosonic networks}

We consider an $M$ mode bosonic network, with squeezed-state inputs
to $N$ out of $M$ modes. A linear, unitary transformation is made
to a set of $M$ output modes, combined with decoherence and losses.
Measurements are carried out on the output state $\hat{\rho}^{\mathrm{\text{out}}}$.
 The theoretical problem is to calculate quadrature correlations
and binned counts for the output quantum state. To solve this, we
utilize discrete Fourier transform methods for ensemble averages of
grouped photon counts \citep{opanchuk2018simulating}.

A squeezed-state is a minimum uncertainty state in which one of the
input mode quadratures has its fluctuations reduced below the vacuum
noise level \citep{yuen1976two,Drummond2004_book,vahlbruch2016detection}.
We suppose that the squeezing of $\hat{\rho}^{\mathrm{in}}$ for each
excited mode is in the imaginary part of $\bm{\alpha}$. If each input
is independent, the quantum state can be factorized into a product
of single-mode states. Defining input quadrature operators $\hat{x}_{j}^{\text{in}}=\hat{a}_{j}^{\text{in}}+\hat{a}_{j}^{\text{in}\dagger}$,
$\hat{p}_{j}^{\text{in}}=\left(\hat{a}_{j}^{\text{in}}-\hat{a}_{j}^{\text{in}\dagger}\right)/i$,
so that $\left[\hat{x}_{\ell}^{\text{in}},\hat{p}_{j}^{\text{in}}\right]=2i\delta_{\ell j}$,
the quantum inputs are defined by a squeezing vector $\bm{r}=\left[r_{1,\ldots}r_{N}\right].$ 

The variances in each mode are:
\begin{align}
\left\langle \left(\Delta\hat{x}_{j}^{\text{in}}\right)^{2}\right\rangle  & =1+\gamma_{j}=e^{2r_{j}}\nonumber \\
\left\langle \left(\Delta\hat{p}_{j}^{\text{in}}\right)^{2}\right\rangle  & =\left(1+\gamma_{j}\right)^{-1}=e^{-2r_{j}}.
\end{align}
The input photon numbers are $n_{j}=\sinh^{2}\left(r_{j}\right)$,
with coherences of $m_{j}=\left\langle \hat{a}_{j}^{2}\right\rangle =\sinh\left(r_{j}\right)\cosh\left(r_{j}\right)$.
Pulsed squeezing \citep{raymer1991limits} involves multiple longitudinal
modes, with mismatches in time or frequency \citep{zhong2020quantum},
as well as phase noise \citep{drummond2020initial}. We model this
experimental decoherence by an intensity transmissivity $T=1-\epsilon$
into the network, combined with a thermal input of $n_{j}^{th}=\epsilon n_{j}$
uncorrelated photons per mode. Our model is similar to thermal squeezing
\citep{Fearn_JModOpt1988}, except with an invariant photon number.

The overall result is that the photon number is unchanged, and the
coherence of each mode is reduced so that $\left\langle \hat{a}_{j}^{2}\right\rangle =\tilde{m}=\left(1-\epsilon\right)m\left(r_{j}\right).$

\subsection{Phase-space representations}

At large $M$, number state expansions of the input state require
exponentially many expansion coefficients to treat this. Instead,
we use phase-space expansions \citep{Hillery_Review_1984_DistributionFunctions}
which allow a probabilistic representation of the input states. Two
common phase-space approaches are used: the Wigner representation
\citep{Wigner_1932}, and the generalized $P$-representation \citep{Drummond_Gardiner_PositivePRep}.
These methods do not assume Gaussianity, and applications to non-Gaussian
photonic networks were already demonstrated \citep{opanchuk2018simulating,opanchuk2019robustness}.

\subsubsection{Positive $P$-representation}

$P$-representations are normally ordered and therefore do not have
any vacuum noise, making them efficient for simulating photo-detection
measurements. The most suitable technique for non-classical photon-counting
measurements is the generalized $P$-representation \citep{Drummond_Gardiner_PositivePRep,DrummondGardinerWalls1981},
which has been applied to other large-scale bosonic simulations \citep{Drummond:2016}.

In this representation, $\hat{\rho}^{\text{in}}$ is expanded over
a subspace of the complex plane defined by
\begin{equation}
\hat{\rho}^{\mathrm{\text{in}}}=Re\int\int P(\bm{\alpha},\bm{\beta})\hat{\Lambda}\left(\bm{\alpha},\bm{\beta}\right)d\mu\left(\bm{\alpha},\bm{\beta}\right)\,.
\end{equation}
The operator basis $\hat{\Lambda}$ is a off-diagonal coherent state
projector onto multimode Glauber ~\citep{Glauber1963_CoherentStates}
coherent states, and $d\mu\left(\bm{\alpha},\bm{\beta}\right)$ is
an integration measure on the $2M$-dimensional complex space of amplitudes
$\bm{\alpha},\bm{\beta}$, which in some cases reduce to simple real
amplitudes. 

For a squeezed input state $\hat{\rho}^{\mathrm{\text{in}}}$, one
obtains a positive P-distribution on a real subspace with $\left(\bm{\alpha},\bm{\beta}\right)=\left(\bm{x},\bm{y}\right)$,
$d\mu\left(\bm{\alpha},\bm{\beta}\right)=d\bm{x}d\bm{y}$. If the
input is $\hat{\rho}_{1S}\equiv\prod_{j}\left|r_{j}\right\rangle \left\langle r_{j}\right|$
, a product of single-mode squeezed state density matrices, the solution
for a squeezed state, based on one-dimensional coherent state expansions
\citep{janszky1990squeezing} is:
\begin{equation}
P\left(\bm{x},\bm{y}\right)=\prod_{j}C_{j}e^{-\left(x_{j}^{2}+y_{j}^{2}\right)\left(\gamma_{j}^{-1}+1/2\right)+x_{j}y_{j}},
\end{equation}
where the normalization constant is $C_{j}=\sqrt{1+\gamma_{j}}/(\pi\gamma_{j}).$

In this approach, normally ordered operator moments are equivalent
to stochastic moments \citep{Drummond_Gardiner_PositivePRep}, so
that $\left\langle \hat{a}_{j_{1}}^{\dagger}\ldots\hat{a}_{j_{n}}\right\rangle =\lim_{S\rightarrow\infty}\left\langle \beta_{j_{1}}\ldots\alpha_{j_{n}}\right\rangle _{P},$
with quantum expectation values denoted $\left\langle \right\rangle $,
and probabilistic averages with $S$ samples denoted $\left\langle \right\rangle _{P}$.

To create input samples for a squeezed state distribution $P(\bm{\alpha},\bm{\beta})$,
one uses real Gaussian noises with $\left\langle w_{i}w_{j}\right\rangle _{P}=\delta_{ij},$
to generate random phase-space samples $\vec{\alpha}=\left[\bm{\alpha},\bm{\beta}\right]=\left[\alpha_{1},\ldots\alpha_{2M}\right]$.
The stochastic model for a pure or thermalized squeezed state, $\left[\bm{\alpha},\bm{\beta}\right]$,
is given by \citep{drummond2020initial}:
\begin{align}
\alpha_{j} & =\delta_{j+}w_{j}+i\delta_{j-}w_{j+M}\nonumber \\
\beta_{j} & =\delta_{j+}w_{j}-i\delta_{j-}w_{j+M}.
\end{align}
The coefficients $\delta_{j\pm}$ must satisfy $\delta_{j\pm}=\sqrt{\left(n_{j}\pm\tilde{m}_{j}\right)/2}$,
which gives real amplitudes for $n_{j}\le\tilde{m}_{j}$, and complex
amplitudes for $n_{j}>\tilde{m}_{j}$.

\subsubsection{Wigner representation}

Other possible phase-space methods include the Wigner \citep{Wigner_1932,Moyal_1949}
and Q-function \citep{Husimi1940}, methods with symmetric and anti-normal
ordering respectively. These give exponentially larger sampling errors
\citep{drummond2020initial} for intensity correlation due to their
extra vacuum noise, while the Glauber P-representation is singular
for squeezed states. These methods have a classical phase-space, in
which $\beta_{j}=\alpha_{j}^{*}$. The Wigner representation is best
for analyzing multipartite entanglement, with data coming from quadrature
measurements \citep{van2003detecting,sperling2013multipartite,teh2014criteria},
which have been carried out at an increasingly large scale \citep{coelho2009three,Chen2014PhysRevLett.112.120505,roslund2014wavelength,armstrong2015multipartite,yoshikawa2016invited}.

For an input quantum density matrix $\hat{\rho}^{\text{in}}$ the
Wigner distribution $W\left(\boldsymbol{\alpha}\right)$ is written
most compactly as \citep{Louisell,Hillery_Review_1984_DistributionFunctions,drummond2014quantum}:
\begin{equation}
W\left(\boldsymbol{\alpha}\right)=\frac{1}{\pi^{2N}}\int d^{2}\mathbf{z}Tr\left[\hat{\rho}^{\text{in}}e^{i\mathbf{z}\cdot\left(\hat{\mathbf{a}}-\boldsymbol{\alpha}\right)+i\mathbf{z}^{*}\cdot\left(\hat{\mathbf{a}}^{\dagger}-\boldsymbol{\alpha}^{*}\right)}\right]
\end{equation}

More generally, $\sigma$-ordered classical bosonic representations
\citep{Cahill_PhysRev1969} are defined using an $\sigma$ parameter
signifying the relative amount of vacuum noise, with $\sigma=0$ for
normal ordering or P-representations and $\sigma=1/2$ for symmetric
ordering or Wigner representations. For pure and thermalized squeezed
state inputs \citep{olsen2009numerical,drummond2020initial}, the
$\sigma$-ordered classical phase-space stochastic amplitude is 
\begin{align}
\alpha_{j} & =\delta_{j+}w_{j}+i\delta_{j-}w_{j+M},\label{eq:Wigner-input}
\end{align}
 where $\delta_{j\pm}$ has the requirement that:
\begin{align}
\delta_{j\pm} & =\sqrt{\left(n_{j}+\sigma\pm\tilde{m}_{j}\right)/2}.
\end{align}

\subsection{Network Transmission}

Once a set of input states is simulated, it can be transformed and
used to sample the output state in any of these representations. An
input density matrix $\hat{\rho}^{\mathrm{\text{in}}}$ is changed
by a linear photonic network to an output density matrix $\hat{\rho}^{\mathrm{\text{out}}}$.
For unitary transformations $\bm{T}$, the phase-space amplitudes
are transformed deterministically, where $\bm{\alpha}'=\bm{T}\bm{\alpha},\,\bm{\beta}'=\bm{T}^{*}\bm{\beta}$,
in all representations. In the generalized P-representation one may
include a non-unitary transmission matrix to take account of losses,
which is equivalent to a master equation \citep{gardiner2004handbook}.

For normal ordering in a linear network, the output density matrix
has a simple form \citep{drummond2016scaling}, including linear couplings
and losses:
\begin{equation}
\hat{\rho}^{\mathrm{\text{out}}}=\Re\int\int P(\bm{\alpha},\bm{\beta})\hat{\Lambda}\left(\bm{T}\bm{\alpha},\bm{T}^{*}\bm{\beta}\right)\mathrm{d}\mu\left(\bm{\alpha},\bm{\beta}\right)\,.\label{eq:Complex-P-2}
\end{equation}
The input distribution $P(\bm{\alpha},\bm{\beta})$ may no longer
be restricted to the real axes if there are input thermal photons
included to model decoherence with $\bm{n}^{th}\ne0$.

For other types of ordering, vacuum noise must be included from the
reservoirs that couple to the system modes, causing decoherence. This
is achieved by noting that for a vacuum state, the input and output
correlations are identical, and for $\sigma$-ordering, $\left\langle \beta_{i}\alpha_{j}\right\rangle =\left\langle \beta_{i}'\alpha_{j}'\right\rangle =\sigma\delta_{ij}.$
It is therefore necessary to add additional vacuum noise if $\sigma>0$
and $\bm{T}$ is non-unitary. This is achieved through defining an
hermitian decoherence matrix, $\bm{D}\equiv\bm{I}-\bm{T}^{\dagger}\bm{T}$,
which has a decomposition $\bm{D}\equiv\bm{U}\bm{\lambda}^{2}\bm{U}^{\dagger}$,
where $\bm{\lambda}$ is diagonal and positive. The matrix square
root is $\bm{B}=\bm{U}\bm{\lambda}\bm{U}^{\dagger},$ and the output
amplitudes are: 
\begin{equation}
\bm{\alpha}'=\bm{T}\bm{\alpha}+\sqrt{\frac{\sigma}{2}}\bm{B}\left(\bm{u}+i\bm{v}\right).\text{}\label{eq:Transmission}
\end{equation}
This ensure that vacuum noise is unchanged.

\section{Quantum measurements\label{sec:Measurements}}

We consider two type of measurements in detail, namely quadrature
detection, and efficient photo-detectors that saturate for more than
one count. 

\subsection{Phase-space representations of measurement operators}

\subsubsection{Quadrature detectors}

For quadrature detection, the quadrature phase amplitudes of each
output mode are $\hat{x}_{i}^{\theta}=\hat{a}_{i}e^{-i\theta}+\hat{a}_{i}^{\dagger}e^{i\theta}$,
with special cases, for $\theta_{i}=0,\pi/2$ of $\hat{x}_{i}$ and
$\hat{p}_{i}$, in a rotating frame. The measurable quadrature correlations
are $\mathcal{C}\left(\bm{m},\bm{\theta}\right)$, where $\bm{m}=\left(m_{1},\dots m_{n}\right)$,
and $\bm{\theta}=\left(\theta_{1},\ldots\theta_{n}\right)$ :
\begin{equation}
\mathcal{C}\left(\bm{m},\bm{\theta}\right)=\left\langle \prod_{j=1}^{n}\left[\hat{x}_{j}^{\theta_{j}}\right]^{m_{j}}\right\rangle .
\end{equation}
This is directly simulated in the Wigner representation. The phase-space
variables are $x_{i}^{\theta}=\alpha_{i}^{\prime}e^{-i\theta}+\alpha_{i}^{\prime\dagger}e^{i\theta},$
and the correlations are calculated by replacing $\hat{x}_{j}^{\theta_{j}}\rightarrow x_{j}^{\theta_{j}}$,
and averaging over the Wigner function.

Hence, one simply has to generate the input amplitudes according to
Eq (\ref{eq:Wigner-input}), transform them according to Eq (\ref{eq:Transmission}),
and average over an ensemble of random events to obtain the output
measured correlations. These are given a detailed analysis in Section
\ref{sec:N-partite-entanglement}.

\subsubsection{Photon number detectors}

For photon number resolving photodetectors, the output number operator
is $\hat{n}'_{j}=\hat{a}_{j}^{\dagger\text{out}}\hat{a}_{j}^{\text{out}}$.
The $n$-th order Glauber correlation function is \citep{Glauber1963_CoherentStates}
:
\begin{equation}
G^{\left(n\right)}\left(c_{j}\right)=\left\langle :\prod_{j=1}^{M}\left(\hat{n}'_{j}\right)^{c_{j}}:\right\rangle ,
\end{equation}
 where $c_{j}=0,1,2\ldots\dots$ is the number of counts at the $j$-th
detector, and $n=\sum c_{j}$ is the total measurement order. The
corresponding phase-space observable is obtained by replacing $\hat{n}'_{j}\rightarrow n'_{j}$,
where $n'_{j}=\alpha'_{j}\beta'_{j}$ is the output number variable,
sampled with probability $P(\bm{\alpha},\bm{\beta})$. To calculate
the output number distribution, there are a number of methods \citep{Hamilton2017PhysRevLett.119.170501},
for transforming these correlations into the observed distributions,
but in this paper we focus on the saturating, or on-off detector,
as recent experiments use this type of measurement.

For saturating photodetectors, all non-zero counts give an identical
output. The ``on-off'' click projection operator is \citep{Sperling2012True}
:
\begin{equation}
\hat{\pi}_{j}\left(c_{j}\right)=:e^{-\hat{n}'_{j}}\left(e^{\hat{n}'_{j}}-1\right)^{c_{j}}:,
\end{equation}
 where $c_{j}=0,1$ is the number of measured counts at the $j$-th
detector, and the output number operator is $\hat{n}'_{j}=\hat{a}_{j}^{\dagger\text{out}}\hat{a}_{j}^{\text{out}}$
. Multi-mode results are given by an $M$-digit binary number $\bm{c}$.
This has $2^{M}$ possible patterns available. For a set $S$ of $M_{S}$
sites, each binary number $\mathbf{c}_{S}$ has an $M_{s}$-order
correlation operator of $\hat{\Pi}_{S}\left(\mathbf{c}_{S}\right)=\bigotimes_{j\in S}\hat{\pi}_{j}\left(c_{j}\right).$
The corresponding expectation, 
\begin{equation}
T\left(\mathbf{c}_{S}\right)=\left\langle \bigotimes_{j\in S}\hat{\pi}_{j}\left(c_{j}\right)\right\rangle ,
\end{equation}
 is the Torontonian function \citep{quesada2018gaussian} for Gaussian
inputs, if the set of sites corresponds to all $M$ output channels.
As this is normally ordered, it has a direct correspondence with a
phase-space function in the positive P-representation. The phase-space
observable is given by replacing $\hat{n}'_{j}\rightarrow n'_{j}$,
where $n'_{j}=\alpha'_{j}\beta'_{j}$ is the output number variable
sampled with probability $P(\bm{\alpha},\bm{\beta})$. 

In all cases, the corresponding $M_{s}$-th order moment is simulated
by replacing $\hat{\pi}_{j}$ with the randomly sampled complex number
$\pi_{j}$. We note that the operators $\hat{\Pi}_{S}\left(\mathbf{c}_{S}\right)$
are projectors. As a result, their expectations are the probabilities
\citep{caves1994quantum} of measuring the count pattern $\mathbf{c}_{S}.$ 

Since there are exponentially many possible count patterns $\mathbf{c}$,
the probability of measuring any individual pattern, $\left\langle \hat{\Pi}_{S}\left(\mathbf{c}_{S}\right)\right\rangle $,
becomes infinitesimal for large $M_{s}$. A direct measurement cannot
obtain all such correlations in less than exponential time. Thus,
it is hard to calculate all probabilities, and it is also hard to
measure them \citep{opanchuk2018simulating,opanchuk2019robustness}.

\begin{figure}
\centering{}\includegraphics[width=1\columnwidth]{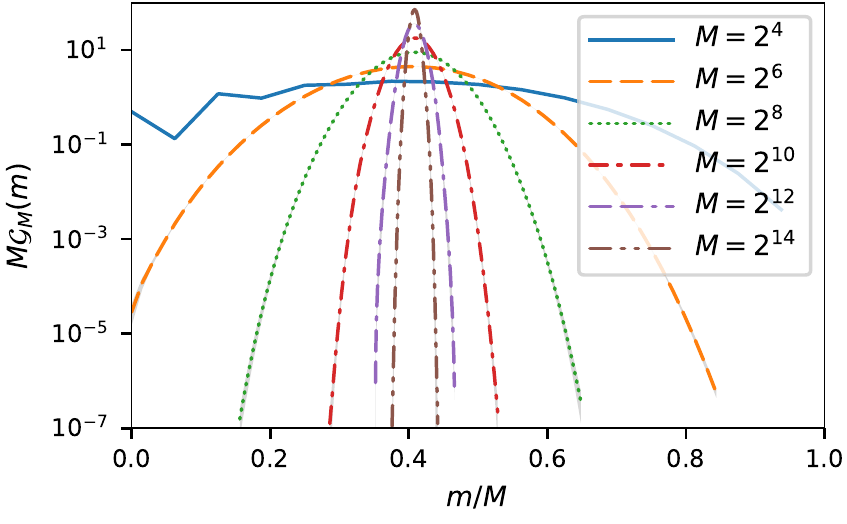}\caption{Theoretical scaling of total grouped count distribution with $M$.
Results are for $M\mathcal{G}_{M}^{(M)}\left(m\right)$ vs $m/M$,
for $\bm{r}=1$, $\epsilon=0$, $N=M/2$ and random unitaries. Mode
numbers were $M=2^{4},2^{6},2^{8},2^{10},2^{12},2^{14}$, with sample
numbers of $10^{8},10^{6},10^{5},1.6\times10^{4},4\times10^{3},2\times10^{3}$
respectively. The transmission matrices are random unitaries. \label{fig:Total-click-distribution-scaling}}
\end{figure}

\subsubsection{Grouped or marginal probabilities}

Grouped counts are therefore essential for verifying GBS statistics
at large $M$, in order to obtain measurable probabilities. One must
simulate and measure the $n$-th order grouped probabilities, $\mathcal{G}_{\bm{S}}^{(n)}\left(\bm{m}\right)$,
where $n=\sum_{j=1}^{d}M_{j}\le M$, is the total probability order
\citep{Glauber1963_CoherentStates}, and:
\begin{equation}
\mathcal{G}_{\bm{S}}^{(n)}\left(\bm{m}\right)=\left\langle {\color{red}{\normalcolor \prod_{j=1}^{d}}}\left[\sum_{\sum c_{i}=m_{j}}\hat{\Pi}_{S_{j}}\left(\bm{c}\right)\right]\right\rangle .
\end{equation}
These are the $d$-dimensional grouped count probabilities of observing
$\bm{m}=\left(m_{1}\ldots m_{d}\right)$ grouped counts in disjoint
sets $\bm{S}=\left(S_{1},S_{2},\ldots\right)$ of $\bm{M}=\left(M_{1},M_{2},\ldots\right)$
output modes. If $n<M$, they include low-order marginal probabilities
often proposed for verification purposes, with $M-n$ outputs ignored.
The first-order correlation with $n=1$, $S=\{j\}$ is the count probability
in the $j$-th channel. Similarly, $n=M$ and $\bm{S}=\left(\{1\},\{2\},\ldots\right)$
gives the Torontonian. For sequential channel groups, the sets $\bm{S}$
are simply denoted by their sizes $\bm{M}$. Using this notation,
$\mathcal{G}_{M}^{(M)}\left(m\right)\equiv\mathcal{G}_{\{1,2,\dots M\}}^{(M)}\left(m\right)$
is the probability for observing $m$ clicks in any pattern, as reported
in recent experiments \citep{zhong2020quantum}. 

We use the terminology of quantum optics \citep{Glauber1963_CoherentStates}
to define the measurement order, $n$, since the binned correlations
involve simultaneous measurements at $n$ different sites. However,
the $\mathcal{G}_{\bm{S}}^{(n)}\left(\bm{m}\right)$ are probabilities
obtained from some or all of the detectors. One can extract moments
from the resulting distributions, and the resulting statistical moments
can also have various orders from $1,\ldots n$. It is important to
distinguish the original measurement order which depends on the number
of modes, from the statistical moment orders that are extracted later. 

Calculating these quantities appears intractable at first: how can
one compute the sum of exponentially many terms, if each is exponentially
hard? Yet, high-order correlations are readily simulated on replacing
the operator $\hat{\pi}_{i}$ by the phase-space variable $\pi_{i}$,
and averaging over the probability $P(\bm{\alpha},\bm{\beta})$. The
summation over grouped correlations is achieved through defining angles
$\theta_{j}=2\pi/\left(M_{j}+1\right)$, with a Fourier observable
$\tilde{\mathcal{G}}$ defined for $k_{j}=0,\dots M_{j}$ , where
$j=1,\ldots d$. The grouped probability is then obtained from a $d$-dimensional
inverse discrete Fourier transform, so that:
\begin{align}
\tilde{\mathcal{G}}_{\bm{S}}^{(n)}\left(\bm{k}\right) & =\left\langle {\color{red}{\normalcolor \prod_{j=1}^{d}}}\bigotimes_{i\in S_{j}}\left(\pi_{i}\left(0\right)+\pi_{i}\left(1\right)e^{-ik_{j}\theta_{j}}\right)\right\rangle _{P}\nonumber \\
\mathcal{G}_{\bm{S}}^{(n)}\left(\bm{m}\right) & =\frac{1}{\prod_{j}\left(M_{j}+1\right)}\sum_{\bm{k}}\tilde{\mathcal{G}}_{\bm{S}}^{(n)}\left(\bm{k}\right)e^{i\sum k_{j}\theta_{j}m_{j}}.
\end{align}

All combinations of terms vanish in the inverse Fourier transform
except those terms with $\bm{m}$ counts. This algorithmic procedure
is highly scalable. To demonstrate this, two simulation codes were
written and tested. Exact Torontonians were simulated for small networks.
Analytically tractable inputs were used to test large networks. Excellent
agreement was found in all cases.

To demonstrate this technique for quantum squeezed inputs, we graph
the grouped count probability in Fig (\ref{fig:Total-click-distribution-scaling})
for sizes up to $M=16,000$, using squeezed states with $r=1$, $\epsilon=0$,
for $N=M/2$ inputs, and random unitaries.

\section{Grouped count verification in GBS experiments}

The grouped probabilities provide a signature of a quantum state.
Clearly, they must be measurable and have a low sampling error. To
validate results, the theoretical sampling error $E_{t}$ must be
less than the experimental sampling error $E_{e}$, where the experimental
sampling error depends on the number of samples used, and scales as
$E_{e}\propto c_{e}S_{e}^{-1/2}$. Experiment and simulations have
similar time-scales for comparable error-bars.

Due to internal averaging, single group measures are less sensitive
to the unitary as $n$ increases, but are very sensitive to decoherence.
Count ``fingerprints'' with more groups are also needed for a complete
test, and one is computed below. Many such measures are available,
both from experimental data and from simulations.

\subsection{Comparisons with a GBS experiment}

To compare theory to experiment, squeezing vectors and transmission
data from a recent $100-$mode Gaussian boson sampling experiment
were obtained \citep{zhong2020quantum} and simulated with $1.2\times10^{6}$
samples. The data was a $50$ mode vector of amplitudes $\bm{r}$,
a $50\times100$ transmission matrix $\bm{T}$, and $5\times10^{7}$
measured click patterns. The experimental data was used to calculate
grouped correlations, and compared to simulations with a standard
chi-squared test \citep{pearson1900x}, using a lower cut-off of 10
counts per bin \citep{Rukhin2010,knuth2014art}. Statistical methods
and tests of the codes are given in the Appendix.

\begin{figure}
\centering{}\includegraphics[width=1\columnwidth]{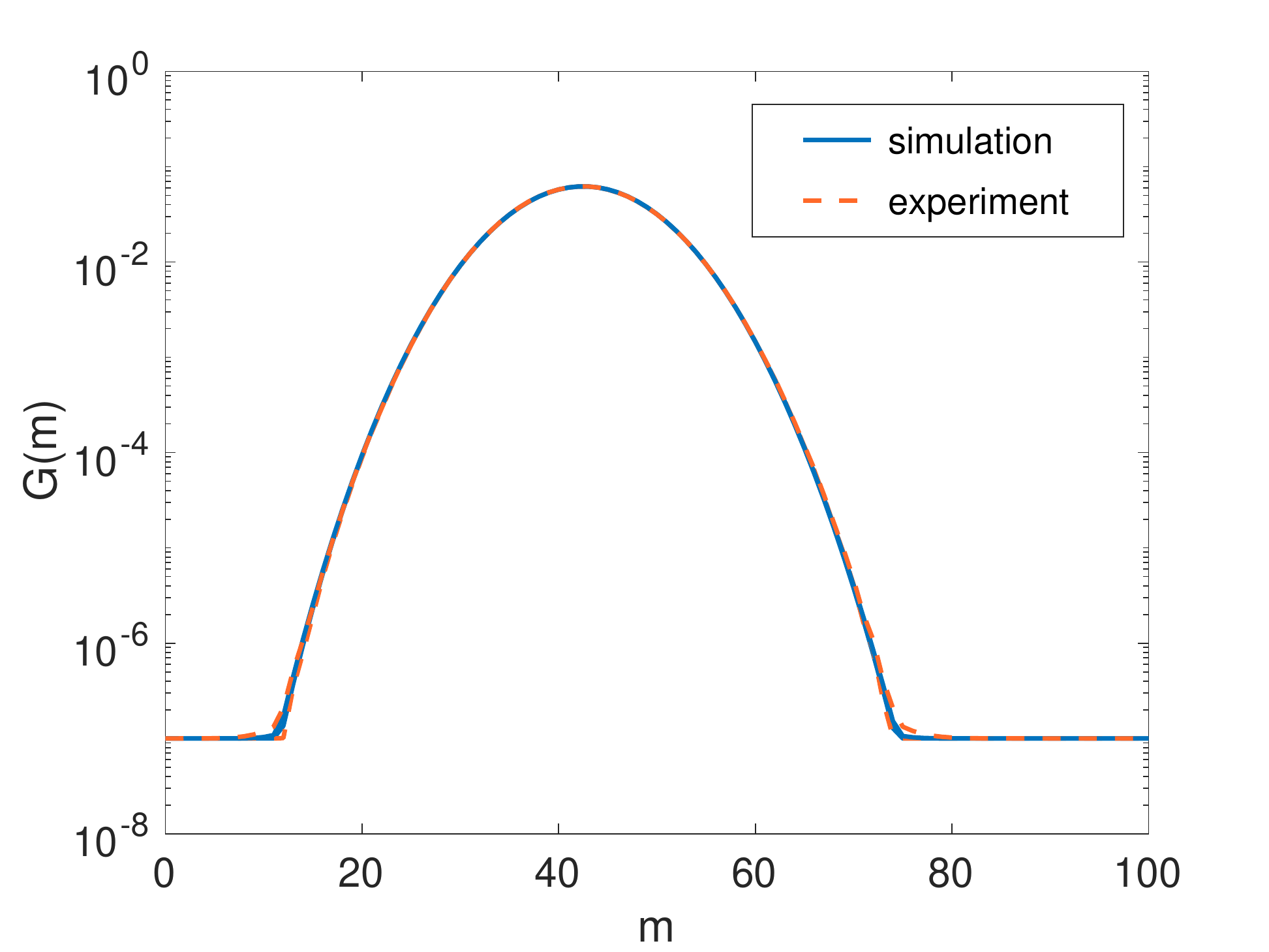}\caption{Comparison of theory with experiment of $\mathcal{G}_{100}^{(100)}\left(m\right)$,
for a $100$ channel GBS total count distribution. Solid blue line
is the theoretical prediction with $\epsilon=0.0932$ relative decoherence
and $1.2\times10^{6}$ samples. The orange dashed line is the experimental
data obtained from $5\times10^{7}$ samples. \label{fig:Comparison-of-theory}}
\end{figure}

\begin{figure}
\centering{}\includegraphics[width=1\columnwidth]{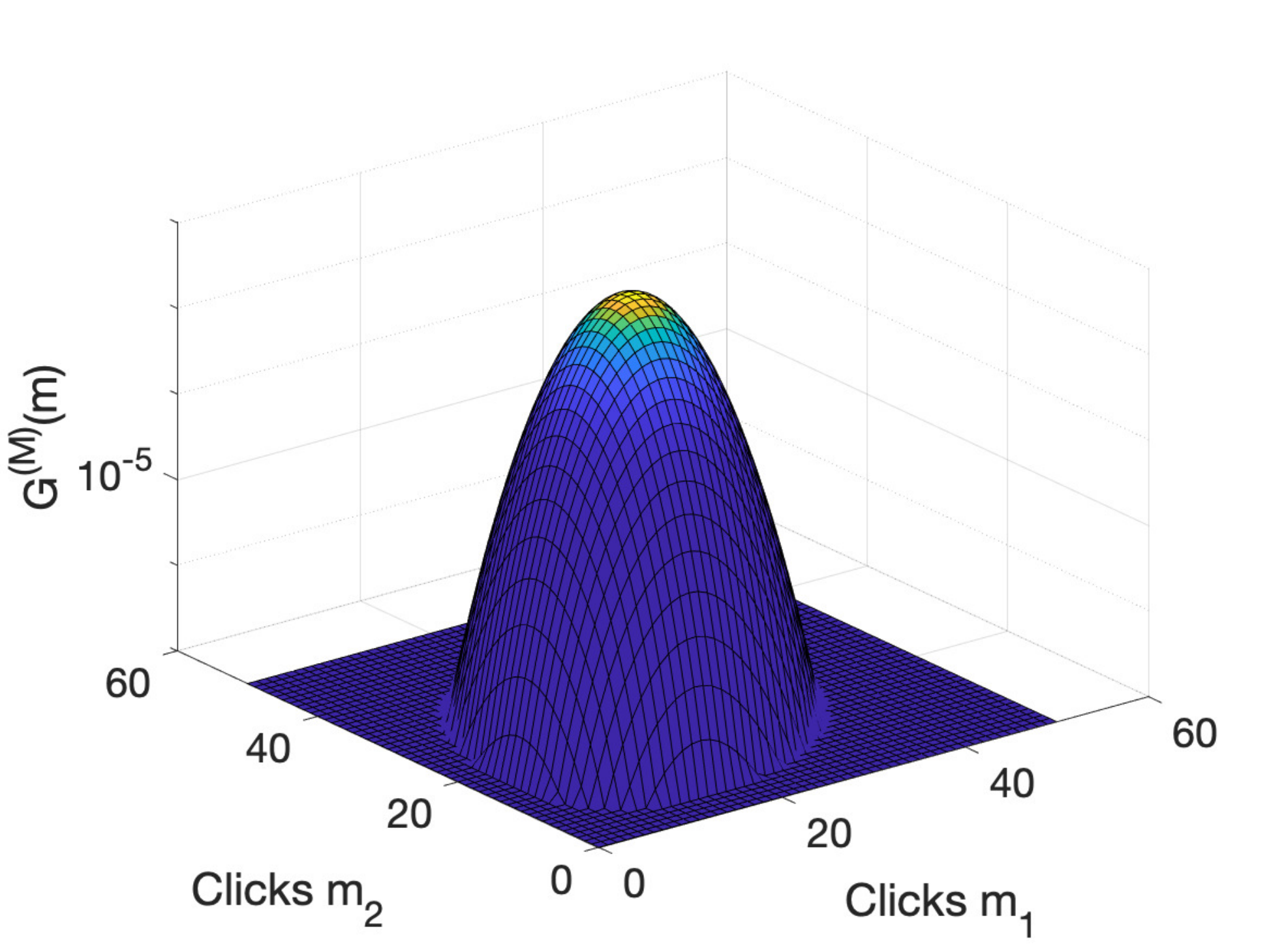}\caption{Simulation of a $100$ channel GBS count distribution binned into
$d=2$ dimensions, $\mathcal{G}_{50,50}^{(100)}\left(m_{1},m_{2}\right),$
with $1.2\times10^{6}$ samples. There are $51^{2}$ data points in
the distribution, leading to over $1000$ distinct data-points. Including
decoherence, the differences between theory and experiment are negligible
on this scale. \label{fig:P2-count distribution}}
\end{figure}
\begin{figure}
\centering{}\includegraphics[width=1\columnwidth]{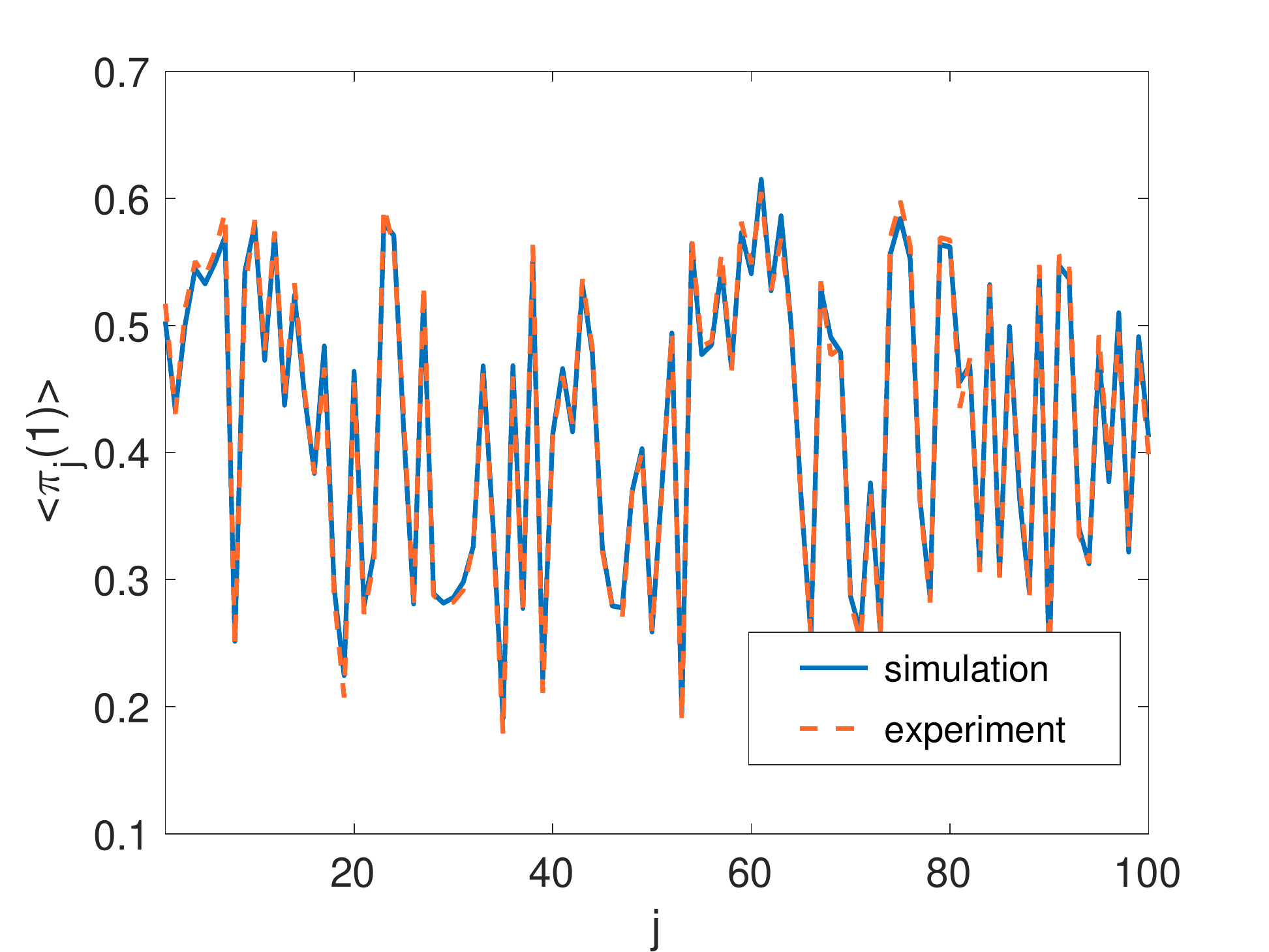}\caption{Comparison of theory vs experiment for a $100$ channel GBS count
probability per channel, $\mathcal{G}_{\{j\}}^{(1)}\left(1\right)\equiv\left\langle \hat{\pi}_{j}\left(1\right)\right\rangle $,
versus mode $j$. Blue line is the theory with with $\epsilon=0.0932$
added decoherence, orange dashed line is obtained from $5\times10^{7}$
experimental data records. Computational sampling errors with $1.2\times10^{6}$
samples were negligible on this scale. \label{fig:Click-probability-per}}
\end{figure}

For $k$ significant bins, one expects $\chi_{c}^{2}/k\approx1$.
Simulating total counts, $\mathcal{G}_{M}^{(M)}\left(m\right)$, with
pure squeezed-state inputs gave a large chi-squared value of $\chi_{c}^{2}/k=9.5\times10^{3}\gg1$,
with $k=63$ valid data points. Additionally, we tested a $100$ mode
fully thermalized model. This gave an even larger discrepancy. The
chi-squared value was $\chi_{th}^{2}/k=6.1\times10^{4}\gg1$, confirming
a prediction \citep{Aaronson:2014} that one can distinguish boson
sampling from uniform distributions.

Better agreement with experiment was obtained with a small admixture
of thermal inputs. For optimal fitting, we included an $\epsilon=0.0932\pm0.0005$
thermal component to model longitudinal mode mismatching. Transmission
amplitudes were multiplied by $1.0235\pm0.0005$. Results of simulations
are given in Fig (\ref{fig:Comparison-of-theory}). This agrees with
experiment over a range of six orders of magnitude in the measurable
grouped probabilities. A chi-squared value of $\chi_{\epsilon}^{2}/k=6.5\pm1$
was obtained, giving three orders of magnitude lower values than with
pure state inputs. Residual discrepancies may be from nonlinearities. 

Fig (\ref{fig:P2-count distribution}) shows $\mathcal{G}_{50,50}^{(100)}\left(m_{1},m_{2}\right)$,
which is a two-dimensional binning of the $100$-th order probabilities.
Any number of bins - up to $M$ - are feasible in principle. However,
experimental sampling errors increase as the grouping dimension increases,
giving a limit of $d=6$ dimensions with currently available experimental
data. As another comparison, the marginal count probability per channel
$\mathcal{G}_{\{j\}}^{(1)}\left(1\right)=\left\langle \pi_{j}\left(1\right)\right\rangle $,
is graphed in Fig (\ref{fig:Click-probability-per}). This also shows
good agreement with experiment. 

\subsection{Detailed statistical comparisons}

We now consider the details of the comparisons and the inferred decoherence
from the grouped $100$-th order correlations in a GBS experiment,
as compared to a phase-space simulation. We wish to compare two hypotheses.
The first, $\mathcal{H}_{0}$, is that the correlations are given
by the experimental squeezing and transmission matrices. The second,
$\mathcal{H}_{1}$, is that there is additional decoherence, modeled
by an thermal fraction $\epsilon$, with an unchanged photon number.

In both cases, the experimental counts are the same. However, graphing
raw experimental and theoretical count probabilities is not useful
for comparative purposes, as the probabilities appear nearly identical
to the naked eye \citep{pearson1900x}. Due to the accuracy of the
data, with over $10^{7}$ total counts, it is much more useful to
graph the normalized deviation between theoretical and experimental
probabilities, as described in the Appendix:
\begin{equation}
z_{m}=\frac{\Delta\mathcal{G}_{M}^{(M)}\left(m\right)}{\sigma_{m}}=\frac{\mathcal{G}\left(m\right)-\mathcal{G}^{e}\left(m\right)}{\sigma_{m}}\,.
\end{equation}

\begin{figure}
\centering{}\includegraphics[width=1\columnwidth]{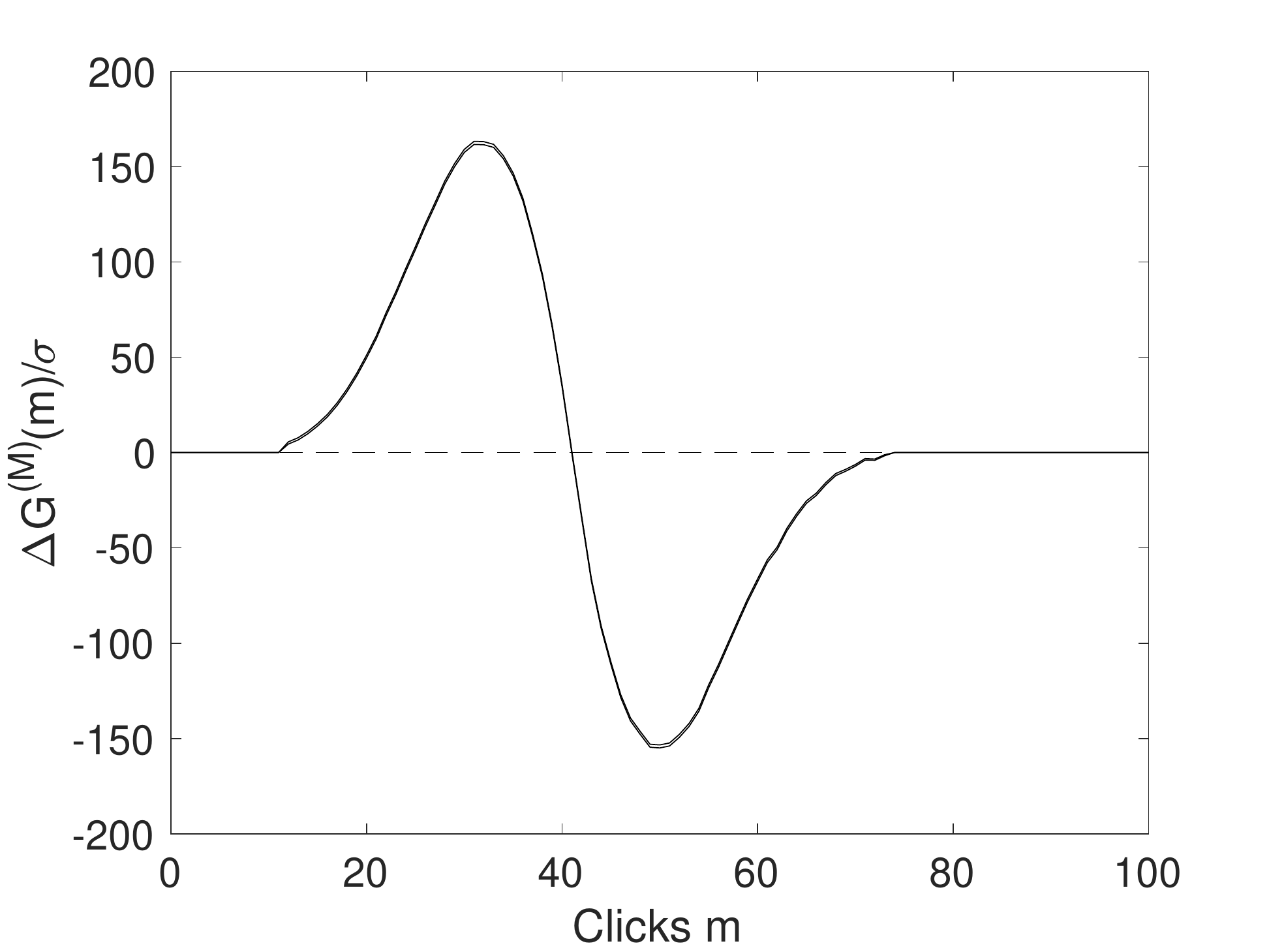}\caption{Normalized difference of simulation versus experimental count distribution,\emph{
excluding} decoherence. Results are for $\Delta\mathcal{G}_{M}^{(M)}\left(m\right)/\sigma_{m}$
vs $m$, with sample numbers of $1.2\times10^{6}$. The error bars
indicate errors due to finite experimental counts plus theoretical
sampling errors. The results are cut off for all counts less than
$10$. \label{fig:Total-click-distribution-scaling-1}}
\end{figure}
For good agreement between theory and experiment, one expects a normalized
difference of unity. Figure (\ref{fig:Total-click-distribution-scaling-1}),
shows the normalized discrepancy between theory and experiment in
$\mathcal{G}_{M}^{(M)}(m)$, in a simulation having no decoherence.
An inspection of the graph shows very significant differences between
the theoretical and experimental count probabilities, with $\left|z\right|\gg1$.
This is reflected in the $\chi^{2}$ results for the null hypothesis
$\mathcal{H}_{0}$, where one obtains an extremely large value of
$5.9\times10^{5}$, out of $63$ valid data points having more than
$10$ counts. This gives a discrepancy ratio of $\chi_{c}^{2}/k=9.5\times10^{3}\gg1$. 

Clearly, when decoherence is excluded, the experiment strongly disagrees
with a simple, coherent GBS model. Therefore, the hypothesis of no
decoherence, apart from losses, has a vanishingly small probability.
The hypothesis of a fully thermal model with $\epsilon=1$ is less
likely still. With this model, the total discrepancy ratio is $\chi_{th}^{2}/k=6.1\times10^{4}\gg1$.
Hence, the output is easily distinguishable from a thermal one \citep{Aaronson:2014}.

\begin{figure}
\centering{}\includegraphics[width=1\columnwidth]{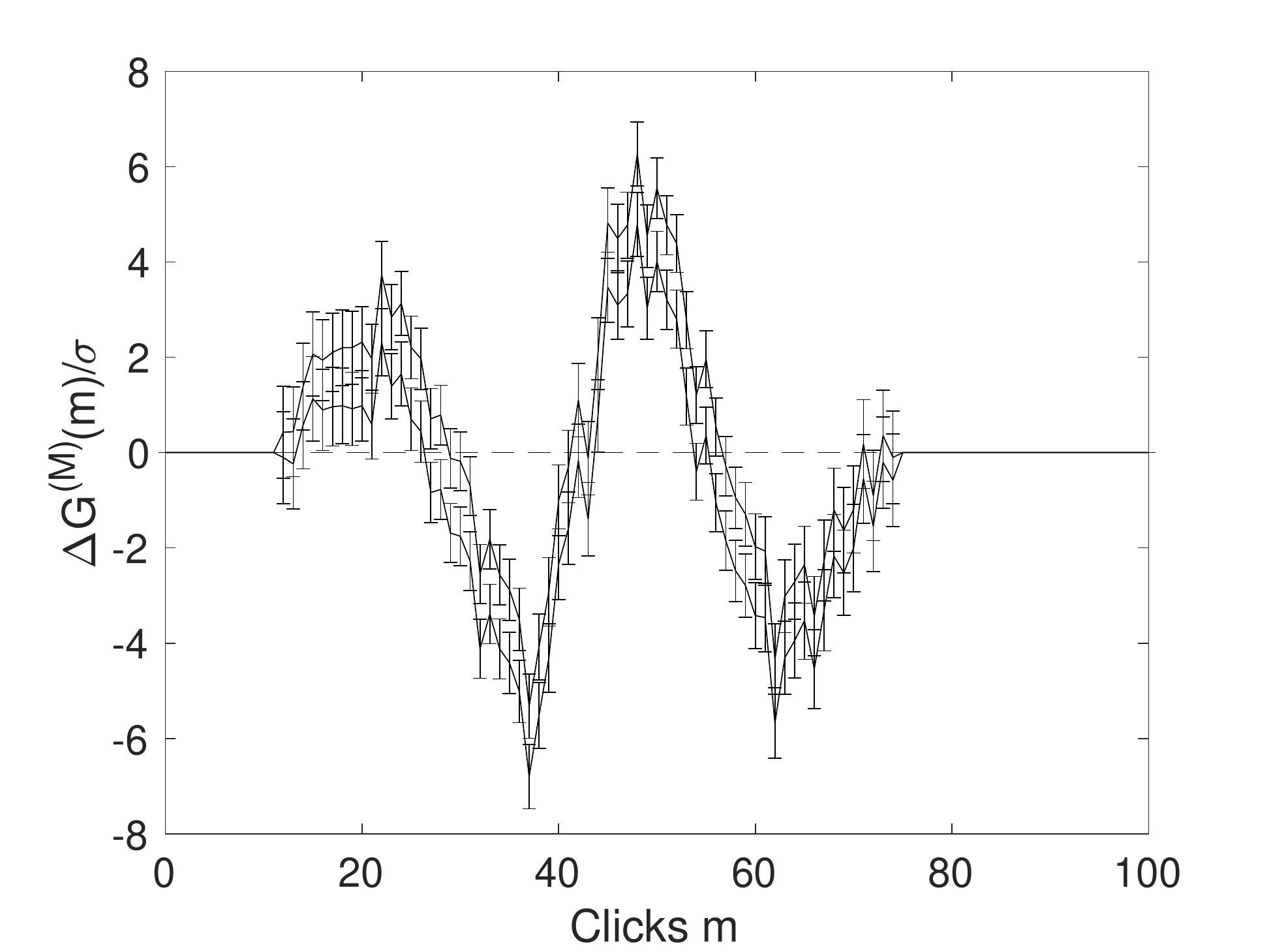}\caption{Normalized difference of simulation versus experimental count distribution,
\emph{including} decoherence. Results are for $\Delta\mathcal{G}_{M}^{(M)}\left(m\right)/\sigma_{m}$
vs $m$, with sample numbers of $1.2\times10^{6}$ . The error bars
indicate RMS errors due to finite experimental counts , plus theoretical
sampling errors (which are 50\% smaller). Results are cut off for
counts less than $10$, where the count data is less reliable. The
maximum error is reduced by about two orders of magnitude compared
to the coherent model. \label{fig:Total-click-distribution-comparisons-with-decoherence}}
\end{figure}

Figure (\ref{fig:Total-click-distribution-comparisons-with-decoherence}),
shows the differences in $\mathcal{G}_{M}^{(M)}(m)$, between experimental
and simulated probabilities with a small decoherence of $\epsilon=0.0932$,
as described above. The transmission amplitude was increased by a
factor of $1.0235$ to improve the fit. This is a small correction,
since even small deviations can result in large chi-squares. The new
graph shows much smaller differences between the theoretical and experimental
count probabilities, with $\left|z\right|\lesssim5$. 

For the hypothesis $\mathcal{H}_{1}$, with additional decoherence,
the $\chi_{\epsilon}^{2}$ value is $400\pm50$. The total ratio is
$\chi_{\epsilon}^{2}/k=6.5\pm1\sim O\left(1\right)$, more than $1000$
times smaller than for a pure state. This indicates that the hypothesis
of additional decoherence is more compatible with experimental measurements.
These results show good agreement with a model of GBS including a
small thermal decoherence. Other physical effects including nonlinearities
may explain the remaining discrepancies.

The phase-space errors here were $50\%$ less than the experimental
errors, and could be reduced further, at the cost of longer computation
time. There were $40$ times more experimental than theoretical samples.
Hence, this simulation is comparable or better than experimental efficiency.
The error ratio depends on the observations. 

The same technique is applicable to any measurable distribution, including
lower order marginal probabilities and multiple partitions. We plotted
an example of a two-dimensional grouped probability distribution in
the previous subsection. This has similar properties, with a distribution
close to the experimental one, and includes thousands of data points. 

In general it is possible to check the GBS hypothesis for any set
of parameters and marginal or grouped probability distributions. Measurable,
grouped probability distributions of GBS experimental data can be
simulated to high accuracy. We have simulated correlation orders from
first up to 16,000th order. The computational time depends on the
complexity of extraction of the binned correlations, and on the error
requirements.

There are many correlation tests possible. Thus, to disprove a classical
mock-up, one could simply use a large, randomly chosen subset of tests
of all orders. Just as with other RNG tests, it is increasingly unlikely
that a range of statistical tests like this can be faked. We conjecture
that the multidimensional grouped probabilities, as in Fig \ref{fig:P2-count distribution},
have the most potential for this due to their polynomially large number
of probability samples.

\section{N-partite entanglement\label{sec:N-partite-entanglement}}

These experiments typically lead to entangled outputs. However, the
entanglement is demonstrated most directly using a different type
of measurement. We will illustrate this for one type of $M$-partite
entangled state that is generated from one or two squeezed vacuum
states. In this section, we briefly outline the known method for generating
such a state \citep{van2003detecting,teh2014criteria}. In short,
the squeezed inputs are first combined across a single beam splitter
to create a two-mode Einstein-Podolsky-Rosen (EPR) entangled state
\citep{Reid:1989}. One of the outputs is then passed through $M-2$
beam splitters \citep{van2003detecting,teh2014criteria}.

\subsection{Multi-mode entanglement theory}

The overall set-up has $M$ inputs $\hat{a}_{j}^{\text{in}}$, where
the first two inputs are orthogonally squeezed vacuum states. In particular,
$\hat{a}_{2}^{\text{in}}$ is a squeezed vacuum input with $\Delta^{2}\hat{x}{}_{2}^{\text{in}}=e^{-2r}$,
and $\hat{a}_{1}^{\text{in}}$ is squeezed vacuum input with $\Delta^{2}\hat{p}{}_{1}^{\text{in}}=e^{-2r}$.
Here $r>0$ is the squeezing parameter. All other inputs are vacuum
states, implying $\Delta^{2}\hat{x}{}_{j}^{\text{in}}=\Delta^{2}\hat{p}{}_{j}^{\text{in}}$.
The inputs are combined across a total of $M-1$ beam splitters. We
use the notation $\Delta^{2}\hat{x}$ to mean the variance of $\hat{x}$
i.e. $\Delta^{2}\hat{x}=(\Delta\hat{x})^{2}=\langle\hat{x}^{2}\rangle-\langle\hat{x}\rangle^{2}$.

To create two-mode EPR entanglement, inputs $1$ and $2$ are passed
through beam splitter $BS1$, with reflectivity $R_{1}^{2}$ and $T_{1}^{2}=1-R_{1}^{2}$,
according to 
\begin{eqnarray}
\hat{a}_{1}^{(1)} & = & R_{1}\hat{a}_{1}^{\text{in}}+T_{1}\hat{a}_{2}^{\text{in}}\nonumber \\
\hat{a}_{2}^{(1)} & = & T_{1}\hat{a}_{1}^{\text{in}}-R_{1}\hat{a}_{2}^{\text{in}}.
\end{eqnarray}
The output of $\hat{a}_{1}$ is $\hat{a}_{1}^{(1)}$. It is straightforward
to show using the approach developed in \citep{Reid:1989} that the
two outputs are EPR correlated with respect to the quadrature phase
amplitudes, i.e. 
\begin{eqnarray}
\Delta^{2}(\hat{x}_{1}-\hat{x}_{2}^{(1)}) & = & 2e^{-2r}\nonumber \\
\Delta^{2}(\hat{p}_{1}+\hat{p}_{2}^{(1)}) & = & 2e^{-2r}.
\end{eqnarray}
More details are given in \citep{Dellios2021,teh2021full}, where
EPR steering is also considered. For large $r$, both variances become
zero. EPR entanglement can also be created from one squeezed input
$\hat{a}_{1}^{\text{in}}$ to give 
\begin{eqnarray}
\Delta^{2}(\hat{x}_{1}-\hat{x}_{2}^{(1)}) & = & 2\nonumber \\
\Delta^{2}(\hat{p}_{1}+\hat{p}_{2}^{(1)}) & = & 2e^{-2r}.
\end{eqnarray}

To generate multipartite entanglement, the field $\hat{a}_{2}^{(1)}$
is combined across a second $BS2$ with reflectivity $R_{2}^{2}$
and $T_{2}^{2}=1-R_{2}^{2}$, according to \citep{van2003detecting}
\begin{eqnarray}
\hat{a}_{2}^{(2)} & = & R_{2}\hat{a}_{2}^{(1)}+T_{2}\hat{a}_{3}^{\text{in}}\nonumber \\
\hat{a}_{3}^{(2)} & = & T_{2}\hat{a}{}_{2}^{(1)}-R_{2}\hat{a}_{3}^{\text{in}}.
\end{eqnarray}
The output of field $\hat{a}_{2}$ is $\hat{a}_{2}^{(2)}$. For $M=3$,
there are only two beam splitters, and the output of $\hat{a}_{3}$
is $\hat{a}_{3}^{(2)}$. For $M=4$, the process continues with another
beam splitter 
\begin{eqnarray}
\hat{a}_{3}^{(3)} & = & (R_{3}\hat{a}_{3}^{(2)}+T_{3}\hat{a}_{4}^{\text{in}})\nonumber \\
\hat{a}_{4}^{(3)} & = & (T_{3}\hat{a}{}_{3}^{(2)}-R_{3}\hat{a}_{4}^{\text{in}}).
\end{eqnarray}
The output of mode $\hat{a}_{3}$ is $\hat{a}_{3}^{(3)}$and the output
of mode $\hat{a}_{4}$ is $\hat{a}_{4}^{(3)}$.

It is possible to continue in this way, and to select the reflectivities
of a string of beam splitters so that we obtain, from two squeezed
inputs, the following solution for the final outputs $\hat{a}{}_{i}^{\text{out}}$,
given by $\hat{a}{}_{i}^{\text{out}}=\hat{a_{i}}^{(i)}$, $i=1,..,M-1$
and $\hat{a}{}_{M}^{\text{out}}=\hat{a_{i}}^{(i-1)}$:
\begin{align}
\xi_{x}=\Delta^{2}(\hat{x}{}_{1}^{\text{out}}-\frac{1}{\sqrt{M-1}}\sum_{i>1}^{M}\hat{x}{}_{i}^{\text{out}}) & =2e^{-2r}\nonumber \\
\xi_{p}=\Delta^{2}(\hat{p}{}_{1}^{\text{out}}+\frac{1}{\sqrt{M-1}}\sum_{i>1}^{M}\hat{p}{}_{i}^{\text{out}}) & =2e^{-2r}.\label{eq:var}
\end{align}
To achieve this, the reflectivity $R_{k}^{2}$ for $k$-th beam splitter,
where $k=1,..,M-1$, is $R_{M-1}^{2}=1/2$, $R_{M-2}^{2}=\frac{1}{3}$,
$R_{M-k}^{2}=1/\left(k+1\right)$ for $k<M-1$, with $R_{1}^{2}=1/2$.
This is explained in more detail in \citep{van2003detecting,teh2014criteria}.

\subsection{Unitary matrix}

The corresponding unitary matrix for the set-up is obtained by first
introducing a vector of reflection amplitudes, defined by
\begin{align}
R_{j} & =\sqrt{\frac{1}{M-j+1}},\,\,\,\,1<j\le M\nonumber \\
R_{1} & =\sqrt{\frac{1}{2}}\nonumber \\
T_{j} & =\sqrt{1-R_{j}^{2}}.
\end{align}
We express the transformation of the $M$ input modes into $M$ genuinely
entangled output modes as a unitary matrix $U$. The output modes
are 
\begin{equation}
\left(\begin{array}{c}
\hat{a}_{1}^{\text{out}}\\
\\
\\
\\
\hat{a}_{k}^{\text{out}}\\
..\\
\\
\hat{a}_{M}^{\text{out}}
\end{array}\right)=U\left(\begin{array}{c}
\hat{a}_{1}^{\text{in}}\\
\\
\\
\\
\hat{a}_{k}^{\text{in}}\\
..\\
\\
\hat{a}_{M}^{\text{in}}
\end{array}\right).
\end{equation}
Defining $R_{0}=-1$ and $R_{M}=1$, the elements $U_{kj}$ of the
$M\times M$ unitary $U$ matrix for $j,k=1,..M$ are given by:
\begin{equation}
\begin{cases}
U_{kj}=0 & j>k+1\\
U_{kj}=R_{1} & j=k=1\\
U_{kk}=-R_{k}R_{k-1}\\
U_{k(k+1)}=T_{k} & k<M\\
U_{kj}=-R_{k}T_{k-1}..T_{j}R_{j-1} & 1\le j<k+1
\end{cases}.
\end{equation}

\subsection{N-partite entanglement simulations}

We now consider how to use phase-space simulations to model an entangled
bosonic network, such as the one described in the last section. This
result can be readily simulated, and we find that one can verify genuine
$M$-partite entanglement \citep{van2003detecting,teh2014criteria},
where we take $M=100$. The optimal simulation method is the Wigner
representation, which for quadrature measurement is the natural approach,
requiring no ordering corrections. Other methods gave larger sampling
errors. As one might guess intuitively, it is optimal to use the representation
that matches the measurement operator \citep{Dirac_RevModPhys_1945}.

This case demonstrates the high efficiency of Wigner phase-space
methods for simulating quadrature measurements, although it is a completely
different type of measurement to the click detection often used in
GBS.

We follow the definitions given in Eq. (\ref{eq:var}). Let $\hat{u}=\hat{x}_{1}-\frac{1}{\sqrt{M-1}}(\hat{x}_{2}+\hat{x}_{3}+..\hat{x}_{M})$
and $\hat{v}=\hat{p}_{1}+\frac{1}{\sqrt{M-1}}(\hat{p}_{2}+\hat{p}_{3}+..\hat{p}_{M})$,
then the observation of 
\begin{equation}
(\Delta\hat{u})(\Delta\hat{v})<\frac{2}{\left(M-1\right)}
\end{equation}
confirms $M$-partite entanglement for all $M$. The observation of
\begin{equation}
(\Delta\hat{u})^{2}+(\Delta\hat{v})^{2}<\frac{4}{M-1}
\end{equation}
also confirms $M$-partite entanglement for all $M$. The proof of
the latter inequality is given in \citep{van2003detecting}, for full
tripartite inseparability. The proofs for genuine $N$-partite entanglement,
and for the first inequality involving a product, follow along similar
lines, using the methods developed in \citep{teh2014criteria}. The
detailed proofs of these threshold points will be given elsewhere
\citep{Dellios2021}. 

The above inequalities suffice to confirm the $M$-partite entanglement
of the fields created by the ideal network, but other methods of detection
are also possible \citep{sperling2013multipartite,gerke2015full,shalm2013three}.
This is particularly true if one assumes pure or Gaussian states,
or is interested to measure full $N$-partite inseparability only
\citep{sperling2013multipartite,gerke2015full,Villar_PRL2005,Chen2014PhysRevLett.112.120505,coelho2009three,shalm2013three}.

\begin{figure}
\centering{}\includegraphics[width=1\columnwidth]{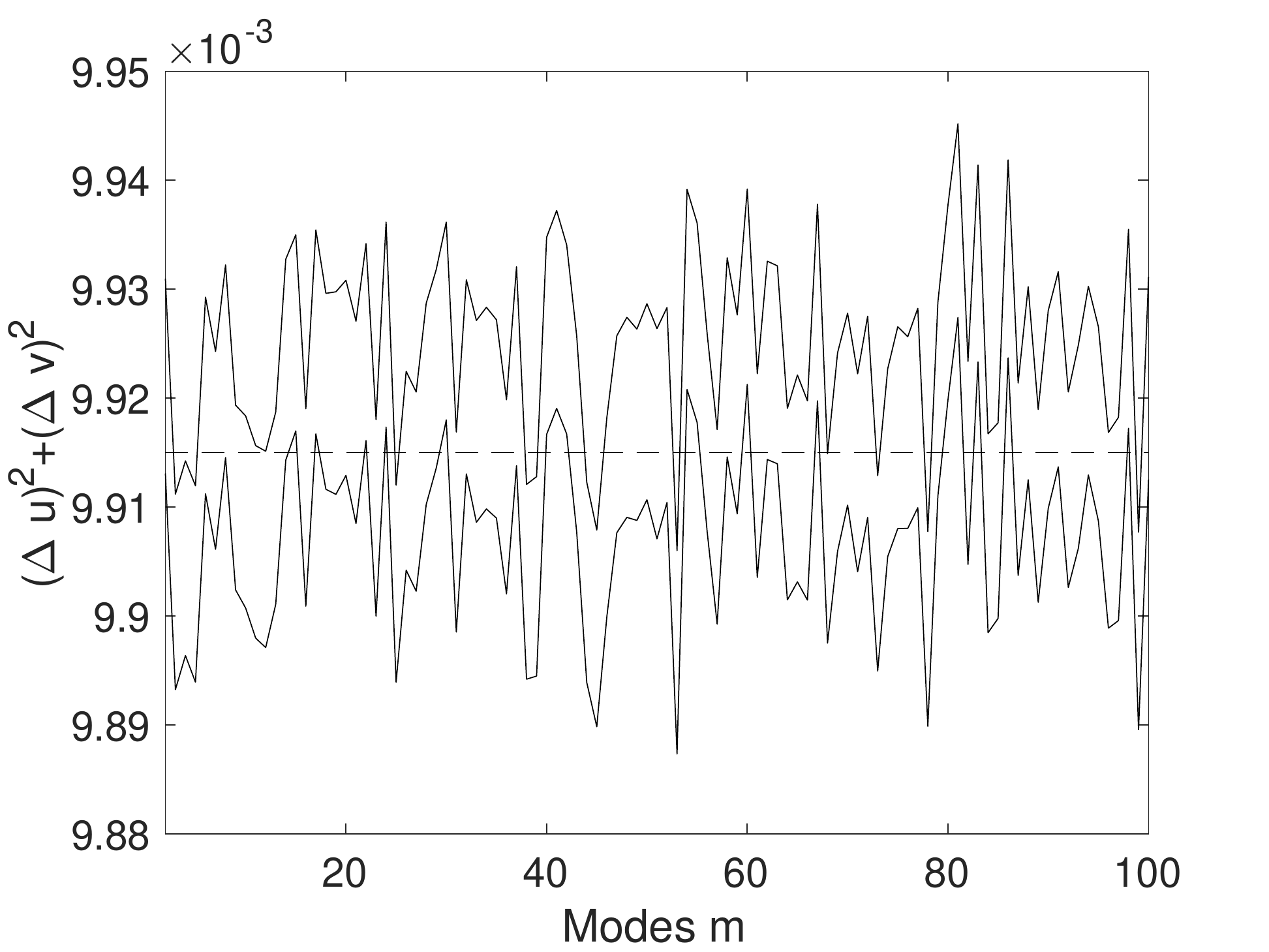}\caption{Graph of simulated multipartite entanglement product against number
of entangled modes, using the Wigner representation. Sample numbers
were $1.2\times10^{6}$ with an input squeezing of $r=3$, using a
unitary matrix and pure state inputs. The upper and lower solid lines
are sampling errors\emph{,} the dashed line the exact result. Sampling
errors here are about $\pm1.0\times10^{-5}$. \label{fig:Variance product-W}}
\end{figure}

Figure (\ref{fig:Variance product-W}), shows the result of a Wigner
simulation of multipartite entanglement, plotted against the number
of input modes, for $r=3$ and $S=1.2\times10^{6}$ samples. The total
ratio of $\chi^{2}/k=0.965<1$, for 99 data points, showing that the
simulation is consistent with the analytic result. The simulation
error bars are $O\left(10^{-5}\right)$. The threshold for the signature
in this case is $0.0404$ at $M=100$, so the criterion is satisfied. 

This level of precision are not obtained for all phase-space methods.
In a positive-P simulation of multipartite entanglement, otherwise
identical to figure (\ref{fig:Variance product-W}) the total chi-squared
ratio was $\chi^{2}/k=0.98<1$, for $99$ independent points, indicating
agreement with the analytic result, but the sampling error bars were
$\pm2\times10^{-3}$, which is $200$ times larger.  Similar large
errors are found for the Q-function. In both cases, one must add or
subtract corrections to transform the variance to symmetric ordering,
which leads to larger sampling errors.

\begin{figure}
\centering{}\includegraphics[width=1\columnwidth]{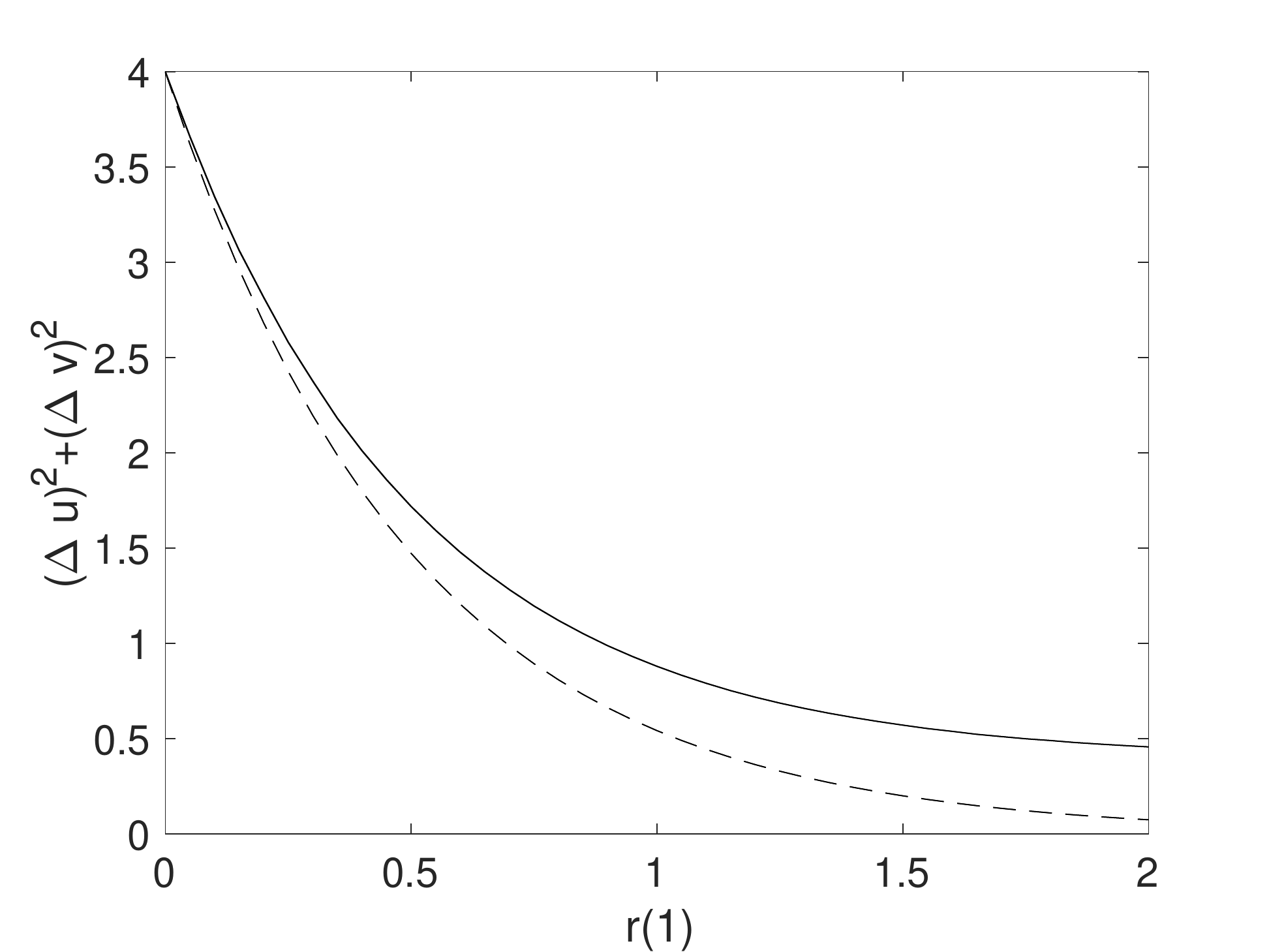}\caption{Graph of simulated multipartite entanglement product in the Wigner
representation, versus squeezing $r$, with an input amplitude transmission
of $t=0.95$ and $M=40$ modes. The required threshold of $0.44$
is not reached even with a large squeezing of $r=2$. Other parameters
as in Fig (\ref{fig:Variance product-W}). Sampling errors are negligible:
$\pm10^{-6}$. This shows that input coupling losses can destroy the
multi-partite entanglement signature. \label{fig:Variance product-loss-40}}
\end{figure}

Phase-space simulations can readily include losses, decoherence and
inhomogeneity. These all impact the amount of input squeezing required
in realistic experiments. A simple example is shown in Fig (\ref{fig:Variance product-loss-40})
which simulates an input coupling amplitude transmission of $0.95$.
This is sufficient to prevent the multipartite signature from being
achieved for an $M=40$ network, with $r=2$. 

\section{Summary}

In summary, we have simulated Gaussian bosonic networks with phase-space
methods. This efficiently simulates large networks with nonclassical
inputs and decoherence. Up to $M=2^{14}=16,384$ modes were treated.
There is excellent agreement with a recent $100$-mode Gaussian boson
sampling experiment for the total count probability, provided thermal
decoherence is included. Other tests, including arbitrary order marginals,
are also possible. The main limitation is that the phase-space sampling
errors can be significant if the grouped probabilities are too small,
but it is straightforward to increase sample numbers to reduce this.
Similar limitations due to sampling error hold for the experimental
data as well, 

More generally, the representation used should be targeted to the
measurement. Positive P-representations are optimal for normally ordered
photo-detectors used in Gaussian boson sampling, while Wigner representations
scale better for quadrature measurements and entanglement. Following
the submission of our work, preprints have appeared covering related
topics \citep{bulmer2021boundary,villalonga2021efficient}, including
calculations of low order marginals of grouped distributions, and
improved direct sampling methods.
\begin{acknowledgments}
PDD thanks Jian-Wei Pan for access to experimental data. Corrected
results were downloaded on Oct 4, 2021. Large-scale calculations were
performed on the OzSTAR national supercomputing facility funded by
Swinburne University of Technology and the Australian National Collaborative
Research Infrastructure Strategy (NCRIS). This work was also funded
through an Australian Research Council Discovery Project Grant DP190101480,
and a grant from NTT Research.
\end{acknowledgments}

\section*{Appendix: Statistics and numerical validation}

\subsection*{Statistical tests}

Statistical tests are essential in comparing theory to experimental
data. In this paper, we compare phase-space simulations both with
exactly known distributions, and with $100$ mode experimental observations.
The test procedures are similar in both cases. We use chi-square methods
originally discovered by Pearson \citep{pearson1900x}, which are
widely used in probability and RNG validation \citep{Rukhin2010,knuth2014art}.
Other tests of probability difference are also feasible, since our
techniques generate complete number distributions, but chi-square
tests are preferable for sampled data because they take account of
experimental sampling errors.

Chi-square tests are used to compare a theoretical probability distribution
to a set of experimental measurements \citep{pearson1900x}, and can
also be used to compare two independent samples. In these tests, experimental
observations are grouped into disjoint classes, with frequencies $f_{i}$
(for $i=1,2,\ldots,k$), from $\mathcal{N}_{e}$ observations. 

Let an hypothesis $\mathcal{H}$ give a probability $P_{i}$ for an
observation in the $i$-th class. Defining an experimental probability
estimate as $P_{i}^{e}=f_{i}/\mathcal{N}_{e}$, $\chi^{2}=\sum_{i=1}^{k}\left(P_{i}^{e}-P_{i}\right)^{2}/\left(\sigma_{e,i}^{2}\right).$
This has a $\chi^{2}$ distribution with $\left\langle \chi^{2}\right\rangle /k=1$,
provided the counts all have a nearly Gaussian distribution.

Here, $\sigma_{e,i}^{2}=P_{i}/\mathcal{N}_{e}\approx f_{i}/\mathcal{N}_{e}^{2}$
is the expected variance in the experimental counts, which have Poissonian
fluctuations. To deal with small counts, it is commonly recommended
that these should not be included if $f_{i}<f_{i}^{min}$. Knuth \citep{knuth2014art}
suggests $f_{i}^{min}=20$, and $f_{i}^{min}=5$ is recommended by
NIST \citep{roscoe1971investigation,Rukhin2010}. We take the middle
ground, ignoring counts less that $f_{i}^{min}=10$. Changing this
threshold has little effect. 

The true theoretical probability $P_{i}$ is not always available.
In this work, we use an estimated value from phase-space simulations,
which converges to $P_{i}$ in the limit of a large ensemble. The
theoretical probability is estimated numerically from its ensemble
mean, $\bar{P}_{i}$ . 

To obtain an error estimate for $\bar{P}_{i}$, it is computed numerically
\citep{opanchuk2018simulating} by using $\mathcal{N}_{s}\gg1$ sub-ensembles,
each with many samples. From the central limit theorem, sub-ensemble
means are nearly Gaussian distributed, with a standard deviation of
$\sigma_{s,i}$. These are obtained from the simulations. As a result,
the simulated ensemble mean $\bar{P}_{i}$ has a standard deviation
in the mean of $\bar{\sigma}_{s,i}=\sigma_{s,i}/\sqrt{\mathcal{N}_{s}}$. 

This uncertainty in the true probability $P_{i}$ implies that the
chi-squared test must be modified, which is similar the well-known
case of two samples drawn from the same population \citep{pearson1911probability,pearson1932probability}.
For a finite ensemble, we employ an error measure of:
\begin{equation}
\chi_{s}^{2}=\sum_{i=1}^{k}z_{i}^{2}=\sum_{i=1}^{k}\frac{\left(\bar{P}_{i}-P_{i}^{e}\right)^{2}}{\sigma_{i}^{2}}.
\end{equation}

This uses the fact that the experimental and simulated data are independent
and nearly Gaussian. The difference in their means has a variance
of $\sigma_{i}^{2}=\sigma_{e,i}^{2}+\bar{\sigma}_{s,i}^{2}$, which
is obtained by adding the two variances. Correlated fluctuations modify
the effective degrees of freedom, so we do not calculate the detailed
$\chi_{s}^{2}$ distribution. However, since $\lim_{\mathcal{N}_{e},\mathcal{N}_{s}\rightarrow\infty}\left\langle \chi_{s}^{2}\right\rangle /k=1$,
we check if $\chi_{s}^{2}/k\sim O\left(1\right)$. 

Fluctuations in the simulated data vanish in the limit of a large
simulation, because $\bar{\sigma}_{s,i}^{2}\rightarrow0$ as $\mathcal{N}_{s}\rightarrow\infty$.
Such tests can be applied to any experimental probability, provided
the measured data is binned to give enough counts to be significant.
This requirement also includes marginal distributions which are included
in our general definition. 

Binned tests are also used in other RNG tests \citep{Rukhin2010},
which have very similar requirements. The difference between our tests
and other RNG tests is that the comparisons are obtained through sampling.
This is necessary because the exact Torontonian is non-computable.
However, it does raise the question of how many samples are needed.
This is answered by increasing the sample number until $\bar{\sigma}_{s,i}<\sigma_{e,i}$.
We found that $1.2\times10^{6}$ was sufficient, using 1200 sub-ensembles
of 1000 samples.

\subsection*{Numerical validation tests}

To test our numerical results, independent numerical codes for simulations
were written for two different languages (Matlab and Python) and computational
platforms (a 14 core desktop, and a supercomputer with GPU hardware).
Simulations were checked against known Torontonians for 16-mode networks
with squeezed inputs \citep{quesada2018gaussian,gupt2020classical}.
The $100$ mode, million sample phase-space simulations took $\sim100s$
on a current desktop computer.

We validated the theoretical code in larger cases by comparison to
exact analytic results for squeezed, thermalized and thermal inputs.
Both unitary and lossy transmission matrices were used, and homogeneous
or inhomogeneous squeezed inputs. For $40\times40$ and $100\times100$
matrices, $9$ different types of moment were tested with up to four-dimensional
binning.

\begin{figure}
\centering{}\includegraphics[width=1\columnwidth]{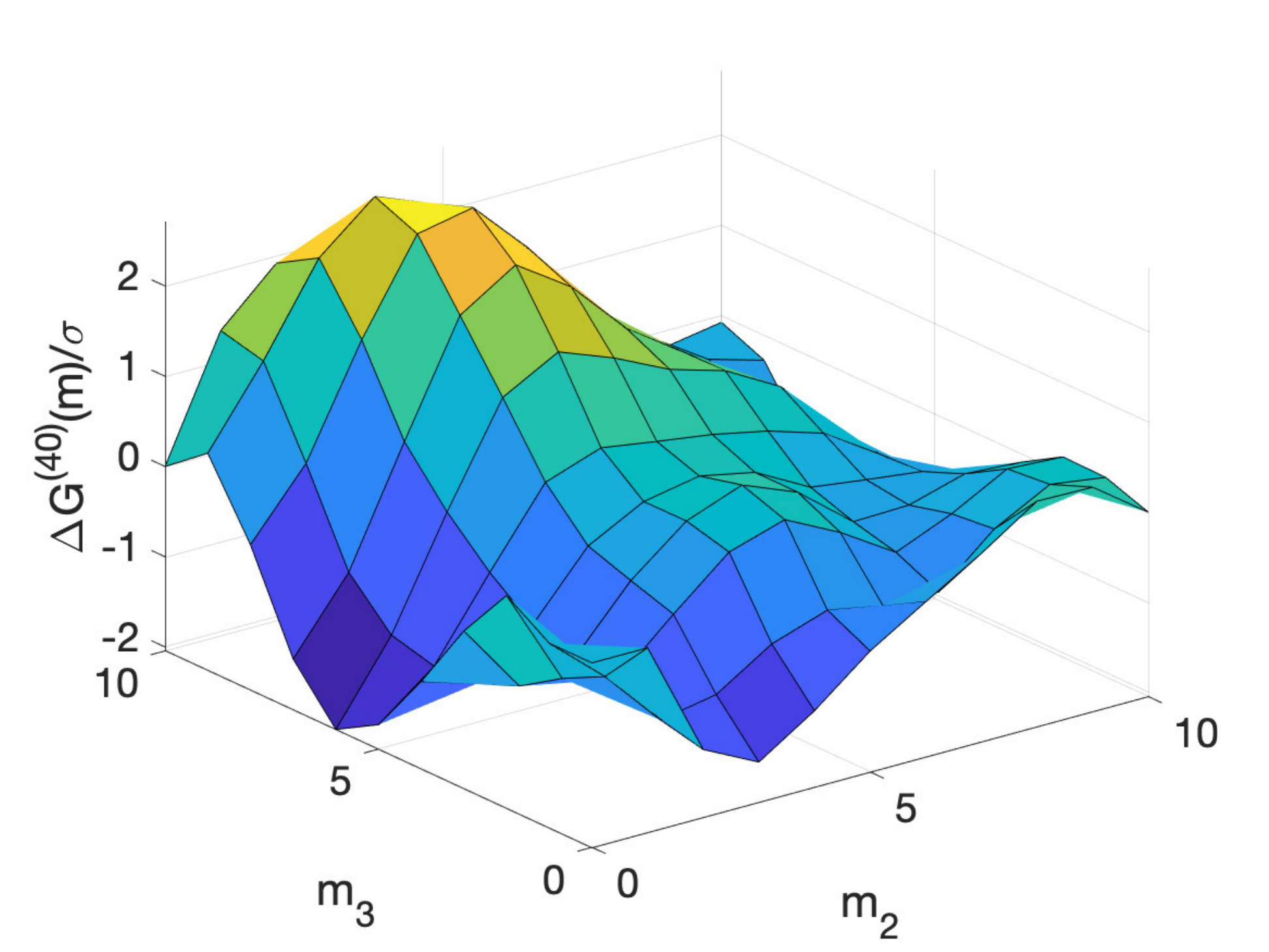}\caption{Normalized difference of simulation versus test distribution, for
a four-fold partition and a thermal input. Results are for $\Delta\mathcal{G}_{(10,10,10,10)}^{(40)}\left(\bm{m}\right)/\sigma_{\bm{m}}$
vs $\bm{m}$, with sample numbers of $1.2\times10^{6}$. Data is given
as a two-dimensional planar slice in $\left(m_{2},m_{3}\right)$ of
a four-dimensional probability space, with $m_{1}=6$ and $m_{4}=5$.
No cut off was required in this slice. \label{fig:Test-multipartite-distribution}}
\end{figure}

A typical example output is plotted in Fig (\ref{fig:Test-multipartite-distribution}),
which shows a test for a thermalized input with $r=\epsilon=1$ and
$n=m=40$, using a random unitary transmission matrix. The output
is the probability for a $40-th$ order correlation, binned four ways,
to give $11^{4}=14641$ click patterns. The graph is a two-dimensional
slice in the $m_{2}-m_{3}$ plane, with $m_{1}=6$ and $m_{4}=5$,
of the normalized error.

Plotted data was all within $\pm2\sigma$. The overall $\chi^{2}$
test gave $\chi^{2}/k=0.99$ in $9935$ significant data points ($P>10^{-7})$,
out of $14641$ possible click patterns. These results show complete
agreement with the analytic probability model.

For each matrix, 68 distinct tests with up to $10^{4}$ data points
were carried out, in P, Q and Wigner phase-space. Probability cutoffs
were used of $\mathcal{G}>10^{-7}$, with $1.2\times10^{6}$ total
samples, since small probabilities are non-Gaussian. This effect is
reduced by increasing the ensemble size. The overall result for $100\times100$
matrices was $\chi_{s}^{2}/k=1.2\pm0.2$. This agrees with analytic
tests, with evidence for nearly Gaussian errors. 

\bibliographystyle{apsrev4-2}

\end{document}